\documentclass[12pt]{article}

\usepackage{amssymb,amsfonts,amsthm}
\usepackage{amsmath}
\usepackage{bm}             
\usepackage[mathscr]{eucal}
\usepackage{cancel}
\usepackage{wasysym}
\usepackage{array}
\usepackage{xcolor}
\usepackage{graphicx}
\usepackage{subfigure}

\def\be{\begin{equation}}
\def\ee{\end{equation}}
\def\bea{\begin{eqnarray}}
\def\eea{\end{eqnarray}}

\def\hsp5{\hspace{5mm}}

\theoremstyle{remark}

\title{\sc Quintessence: Quadratic potentials}

\begin{document}

\author{
\sc Artur Alho,$^{1}$\thanks{Electronic address:{\tt
aalho@math.ist.utl.pt}}\,, Claes Uggla$^{2}$\thanks{Electronic
address:{\tt claes.uggla@kau.se}}\\
$^{1}${\small\em Center for Mathematical Analysis, Geometry and Dynamical Systems,}\\
{\small\em Instituto Superior T\'ecnico, Universidade de Lisboa,}\\
{\small\em Av. Rovisco Pais, 1049-001 Lisboa, Portugal.}\\
$^{2}${\small\em Department of Physics, Karlstad University,}\\
{\small\em S-65188 Karlstad, Sweden.}}

\date{\normalsize{November 15, 2025}}
\maketitle

\begin{abstract}

Arguably one can use a canonical scalar field $\varphi$,
minimally coupled to gravity, with quadratic potentials
$V = \Lambda \pm \frac12 m^2\varphi^2$ to explore some general 
features of slow-roll and hilltop thawing quintessence, respectively. 
For each of these two potentials, and pressure-free matter, 
we introduce a regular unconstrained dynamical system on a compact state 
space, 
where the formulation for the hilltop case is new. Together with a derivation 
of monotonic functions in the two global state space settings, this enables 
us to obtain global results and to introduce figures that illustrate the global 
solution spaces of these models, in which we situate the observationally 
viable quintessence solutions.

\end{abstract}

\newpage

\section{Introduction\label{sec:intro}}

\begin{figure}[ht!]
	\begin{center}
     \subfigure[$V/\Lambda = 1 + \left(\frac{m^2}{2\Lambda}\right)\varphi^2$]{\label{V1}
     \includegraphics[width=0.22\textwidth]{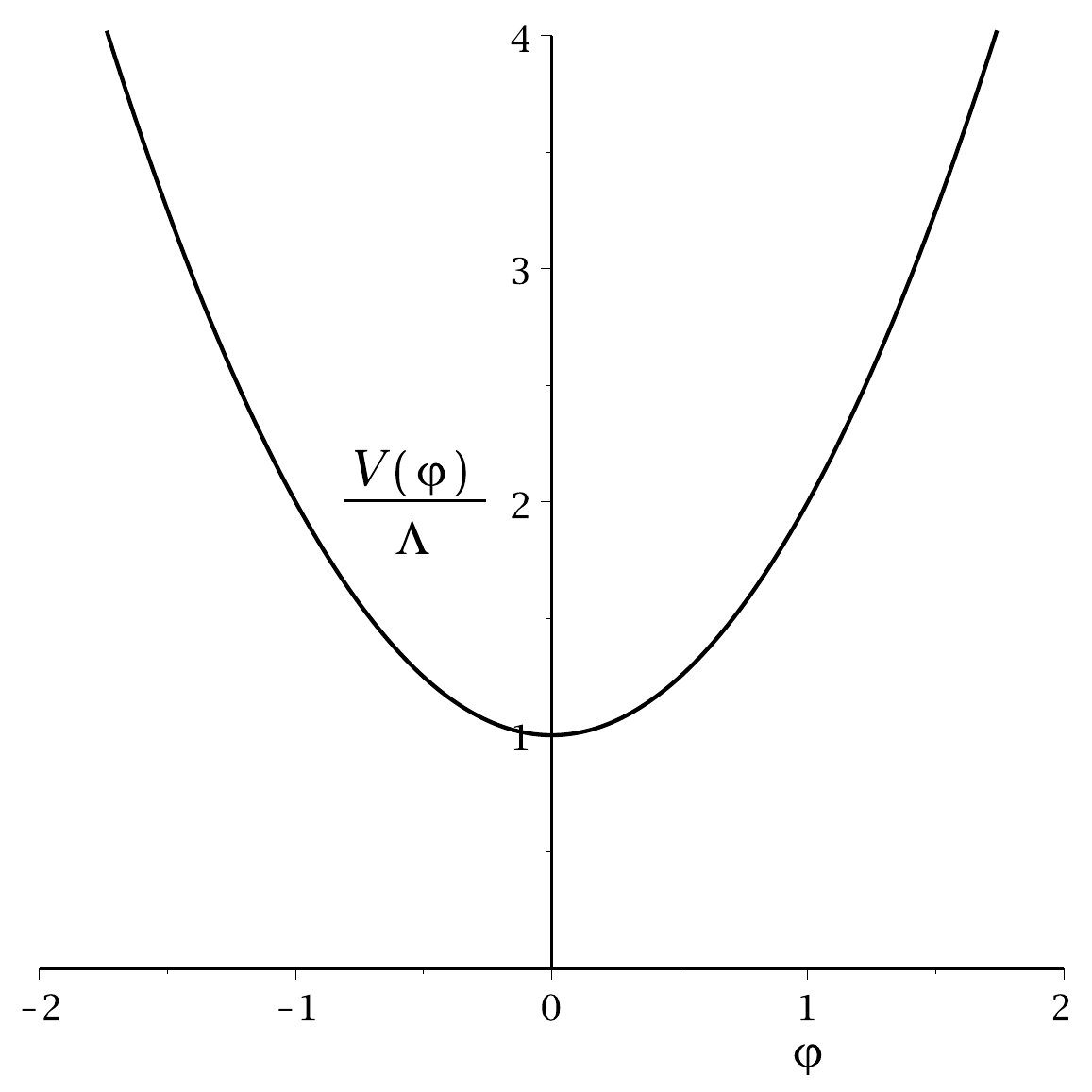}}
		\hspace{2.0cm}
	 \subfigure[$V/\Lambda = 1 - \left(\frac{m^2}{´2\Lambda}\right)\varphi^2$]{\label{V2}
	 \includegraphics[width=0.22\textwidth]{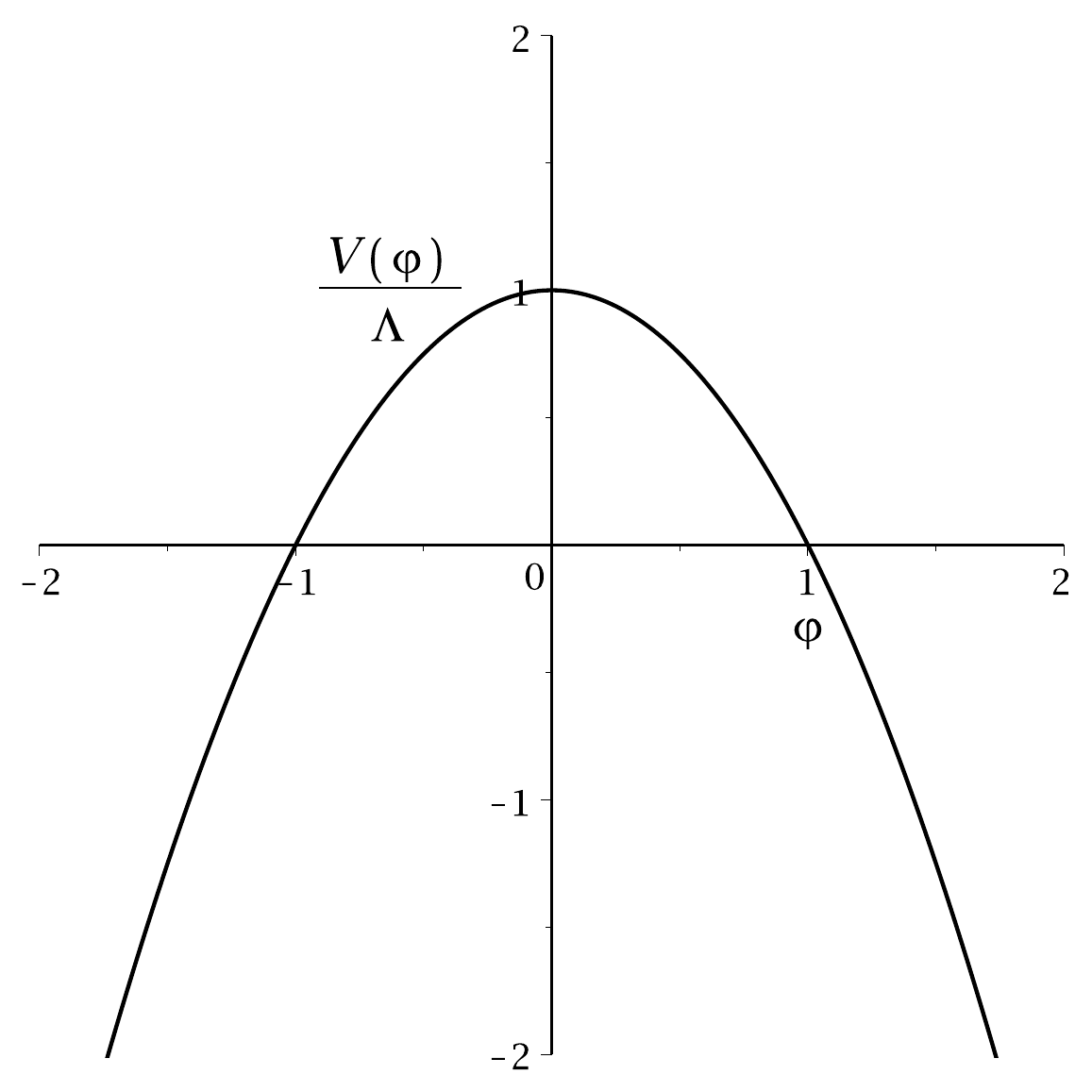}}
			\vspace{-0.5cm}
	\end{center}
\caption{The potentials $V = \Lambda \pm \frac12 m^2\varphi^2$.}
\label{fig:Pot}
\end{figure}

Over the last $\sim 30$ years the $\Lambda$CDM model with a dark energy (DE) equation of state
parameter $w_\mathrm{DE} = w_\Lambda =-1$ has been the standard model in cosmology.
Recent observations, however, such as the Baryonic Acoustic Oscillations (BAO) by the
DESI collaboration~\cite{desicol2025}, suggest that $w_\mathrm{DE}$ might be evolving.
A simple and theoretically appealing dynamical DE component is a canonical scalar
field, $\varphi$, minimally coupled to gravity, generally referred to as quintessence.
By Taylor expanding a potential and redefining the scalar field by an appropriate
translation to complete the square one generically obtains
$V(\varphi) \approx \Lambda \pm \frac12 m^2\varphi^2$. In, e.g., \cite{feletal02} and~\cite{wolf2025}
it was therefore argued that such potentials describe generic
observational features of quintessence.
For this reason we will introduce
\emph{regular unconstrained dynamical systems on compact state spaces}
to describe the global dynamics of quintessence models for which
\begin{equation}
V(\varphi) = \Lambda \pm \frac12 m^2\varphi^2,\qquad \Lambda > 0,
\end{equation}
where $\Lambda>0$ is required in order to have the $\Lambda$CDM solution
as a special case, see Figure~\ref{fig:Pot}.
Note that these potentials are invariant under the transformation
$\varphi \mapsto - \varphi$, which results in a discrete symmetry in the
dynamical systems formulations below.

We further restrict considerations to spatially homogeneous, isotropic and flat
spacetimes with a source that apart from the scalar field $\varphi$ also
consists of matter with an energy density $\rho_\mathrm{m}>0$ and no pressure
$p_\mathrm{m}=0$, representing, primarily, cold dark matter (CDM).\footnote{This
matter model is useful for describing the transition from an epoch of
matter-domination to one in which the scalar field is dominant. A more
realistic model involves both matter and radiation, where an epoch of
radiation-domination precedes that of matter-domination. Although
generalizing the discussion to include radiation is straightforward, it leads
to four-dimensional state spaces instead of the present three-dimensional ones.
For simplicity and pedagogically illustrative
reasons we therefore neglect radiation in this paper.}
This leads to that the general relativistic Friedmann and Raychaudhuri equations;
the (non-linear) Klein-Gordon equation, and energy conservation
for matter with $p_\mathrm{m}=0$, can be written as\footnote{We use units such
that $c=1$ and $8\pi G=1$, where $c$ is the speed of light and
$G$ is the gravitational constant.}
\begin{subequations}\label{Mainsysdim}
\begin{align}
3H^2 &= \rho, \label{Gauss}\\
\dot{H} + H^2 &= -\frac16(\rho + 3p), \label{Ray}\\
\ddot{\varphi} &=-3H\dot{\varphi} - V_{,\varphi}, \label{KG}\\
\dot{\rho}_\mathrm{m} &= -3H\rho_\mathrm{m},
\end{align}
\end{subequations}
where an overdot represents the derivative with respect to the cosmic time $t$;
the Hubble variable is defined by $H=\dot{a}/a$, where $a(t)$ is the cosmological
scale factor, and where the total energy density $\rho$ and pressure $p$
are given by
\begin{subequations}\label{rhop}
\begin{alignat}{2}\label{rhoptot}
\rho &= \rho_\varphi + \rho_\mathrm{m},&\qquad p &= p_\varphi,\\
\rho_\varphi &= \frac12\dot{\varphi}^2 + V(\varphi),&\qquad
p_\varphi &= \frac12\dot{\varphi}^2 - V(\varphi).\label{rhophipphi}
\end{alignat}
\end{subequations}
%
Since~\eqref{Gauss}, \eqref{Ray} and~\eqref{rhop} yield
\begin{equation}\label{Hdotmon}
\dot{H} = -\frac12(\rho + p) = -\frac12\left(\rho_\mathrm{m} + \dot{\varphi}^2\right) < 0,
\end{equation}
it follows that $H$ is monotonically decreasing. For positive scalar field potentials
equation~\eqref{Gauss} implies that $H$ cannot change sign since $\rho>0$ in this
case and as a consequence there are two disjoint state space components:
one with $H<0$ and one forming the cosmologically relevant $H>0$ state space.
This is no longer the case when the scalar field potential can change sign from
positive to negative, as for $V = \Lambda - \frac12 m^2\varphi^2$, since this implies that
$\rho$ can change sign, which results in a single connected state space.

In this paper we introduce regular dynamical systems on three-dimensional compact
state spaces and use dynamical systems methods to study the global solution space
properties, with a focus on quintessence `attractor' solutions. The next section deals
with the case $V = \Lambda + \frac12 m^2\,\varphi^2$ while the subsequent one
addresses $V = \Lambda - \frac12 m^2\,\varphi^2$.

\section{Models with $V=\Lambda + \frac12 m^2\varphi^2$\label{sec:Lp}}

\subsection{Dynamical systems derivation\label{subsec:dynsysderLp}}

We follow~\cite{alhetal23} and define the variables $u$
and $v$ according to
\begin{equation}\label{uvdef}
u = \frac{\varphi^\prime}{\sqrt{3\Omega_\varphi}}, \qquad
v = \sqrt{\frac{\Omega_\varphi}{3}},
\end{equation}
where $\Omega_\varphi$ is the Hubble-normalized scalar field energy density, i.e.,
\begin{equation}
\Omega_\varphi = \frac{\rho_\varphi}{3H^2},
\end{equation}
while the ${}^\prime$ refers to the derivative w.r.t. the $e$-fold time
\begin{equation}\label{Ndef}
N = \ln(a/a_0),
\end{equation}
where subscripts $_0$ henceforth refer to the present time $t=t_0$ and thus
$a_0 = a(t_0)$ and  $t=t_0 \Rightarrow N=0$. The
definition~\eqref{Ndef} implies that $N\rightarrow - \infty$
and $N\rightarrow + \infty$ when $a\rightarrow 0$ and
$a \rightarrow\infty$, respectively. It follows that
\begin{equation}
\varphi^\prime = 3uv,\qquad \Omega_\varphi =3v^2,\qquad w_\varphi = \frac{p_\varphi}{\rho_\varphi} = u^2 - 1,
\end{equation}
and
\begin{equation}
-\sqrt2\leq u\leq\sqrt2, \qquad
0\leq  v\leq  1/\sqrt3.
\end{equation}
Next we introduce a new bounded scalar field variable $\bar{\varphi}$
according to
\begin{equation}\label{barvarphidef}
\bar{\varphi} = 2\arctan\left(\bar{m}\,\varphi\right),\qquad
\varphi = \bar{m}^{-1}\tan\left(\frac{\bar{\varphi}}{2}\right),
\end{equation}
where $\bar{\varphi} \in [-\pi,\pi]$ and $\bar{m}$ is defined as
\begin{equation}
\bar{m} = \frac{m}{\sqrt{2\Lambda}}.
\end{equation}
The definitions~\eqref{uvdef}, \eqref{barvarphidef} and the equations
in~\eqref{Mainsysdim} or, alternatively, the general equations in~\cite{alhetal23}
for bounded
\begin{equation}
\lambda = -\frac{V_{,\varphi}}{V},
\end{equation}
where in the present case
\begin{equation}
\lambda = -\bar{m}\sin\bar{\varphi},\qquad \frac{d\bar{\varphi}}{d\varphi} = \bar{m}(1 + \cos\bar{\varphi}),
\end{equation}
lead to the following three-dimensional unconstrained \emph{regular} dynamical system
for $(\bar{\varphi}, u, v)$:
\begin{subequations}\label{Dynsys.uv(i)}
\begin{align}
{\bar\varphi}^\prime &= 3\bar{m}(1 + \cos\bar{\varphi})uv,\label{barvarphiprimei}\\
u^\prime &= -\frac{3}{2}(2-u^2)\left(u + \bar{m}v\sin\bar{\varphi}\right), \label{u.primei} \\
v^\prime &= \frac{3}{2}(1-u^2)(1-3v^2)v. \label{v.primei}
\end{align}
\end{subequations}

Due to the regularity of the equations, we can include the boundaries, which lead to
a compact state space that is a rectangular box. It is easy to visualize $w_\varphi$
and $\Omega_\varphi$ since $w_\varphi$ is constant on the vertical planes
$u=\mathrm{const.}$ while $\Omega_\varphi$ is constant on the horizontal planes
$v=\mathrm{const.}$. The boundaries are thereby given by a base $v=0$ and top
$v=1/\sqrt3$, representing $\Omega_\varphi=0, 1$, respectively; the sides
$u=\pm\sqrt2$ correspond to $w_\varphi=1$ (since
$\Omega_V = \frac{V}{3H^2} = \frac{3}{2}(2-u^2)v^2$, it follows that $\Omega_V=0$
on these boundaries); the sides $\bar{\varphi} = \pm \pi$
yield equations that are identical to those obtained from a constant potential
$V = \Lambda>0$. Apart from the boundary $v = 1/\sqrt{3}$ $(\Omega_\varphi = 1)$,
the equations on the boundaries are all easily solved, see~\cite{alhetal23}.
Note further that the state space can be viewed as being divided
into a central slab $-1<u<1$, in which $w_\varphi<0$ where thereby $v$ and
$\Omega_\varphi$ are increasing, and two outer slabs $1<u^2<2 $ in which
$w_\varphi$ satisfies $0<w_\varphi<1$ and where $v$ and $\Omega_\varphi$
are decreasing, as follows from~\eqref{v.primei}.

\subsection{Dynamical systems analysis\label{subec:dynsysanalysismp}}

Due to the symmetry $\varphi \mapsto - \varphi$ of the potential the
dynamical system~\eqref{Dynsys.uv(i)} admits the discrete symmetry
\begin{equation}
(\bar{\varphi},u) \mapsto -(\bar{\varphi},u),
\end{equation}
which leads to that orbits (i.e., solution trajectories) occur in pairs related
by this symmetry. The global dynamics of the present models is further restricted
by the monotonic function:
\begin{subequations}\label{3H2mon}
\begin{align}
3H^2 &= \frac{V(\bar{\varphi})}{\Omega_V} = \frac{2V(\bar{\varphi})}{3(2-u^2)v^2},\\
(3H^2)^\prime &= - 2(1+q)(3H^2) = -3(1 - 3v^2 + 3(uv)^2)(3H^2),
\end{align}
\end{subequations}
where
\begin{equation}
q = -\frac{a\ddot{a}}{\dot{a}^2} = -(1 + H^\prime/H) = \frac12[1+9(u^2-1)v^2]
\end{equation}
is the deceleration parameter; since $3H^2 = 2V(\bar{\varphi})/3(2-u^2)v^2$ is
strictly monotonically decreasing in the interior state space $(\bar{\varphi},u,v)$
due to that $1 + q$ and $1 - 3v^2 + 3(uv)^2$ then are positive, it follows that
there are no interior fixed points or recurring orbits in the interior state
space --- all interior orbits begin and end on the boundaries. Note further that
an orbit is accelerating $(q<0)$ when it is inside a trough given by the
parabola-like profile $9(1-u^2)v^2 = 1$, see~\cite{alhetal23} for figures and
further details.

The form of the orbits on the invariant boundaries $v=0$ and $u=\pm\sqrt2$
is independent of the potential. First,
the base of the state space, $v=0$, forms the Friedmann-Lema\^{i}tre
($\mathrm{FL}$) invariant boundary set with $\Omega_\varphi=0$, $\Omega_\mathrm{m}=1$.
Since $\bar\varphi' = 0$ and $u^\prime = -\frac{3}{2}(2-u^2)u$ on this set
it follows that there are three lines of fixed points:
\begin{subequations}\label{FL.fixed}
\begin{alignat}{2}
\mathrm{FL}_0^{\varphi_*}&\!: &\qquad (\bar\varphi,u,v) &= (\bar{\varphi}_*,0,0),\\
\mathrm{FL}_{\pm}^{\varphi_*}&\!: &\qquad (\bar\varphi,u,v) &= (\bar{\varphi}_*,\pm\sqrt2,0),
\end{alignat}
\end{subequations}
with the constant $\bar{\varphi}_*$ satisfying $-\pi\leq \bar{\varphi}_*\leq \pi$,
where $w_\varphi=-1$ for $\mathrm{FL}_0^{\varphi_*}$ and
$w_\varphi=1$ for $\mathrm{FL}_{\pm}^{\varphi_*}$.
These lines of fixed points are connected by
heteroclinic orbits\footnote{A heteroclinic orbit
is an orbit that connects two different fixed points.}
$\mathrm{FL}_\pm^{\varphi_*} \rightarrow\mathrm{FL}_0^{\varphi_*}$ that are straight lines
with $\bar{\varphi} = \bar{\varphi}_* = \mathrm{constant}$, and thus the evolution of the
scalar field is `frozen' on $v=0$.

Second, the orbits on the $u=\pm\sqrt2$ boundaries can also be determined explicitly,
as shown in~\cite{alhugg23,alhetal22}. In particular, all the orbits on the $u=\sqrt{2}$
($u=-\sqrt{2}$) boundary originate from the kinetic scalar field dominated `kinaton'
fixed point $\mathrm{K}^-_+$ ($\mathrm{K}_-^+$) 
at $(\bar{\varphi},u,v)=(-\pi,\sqrt{2},1/\sqrt{3})$ ($(\bar{\varphi},u,v)=(\pi,-\sqrt{2},1/\sqrt{3})$)
and end at the line $\mathrm{FL}_+^{\varphi_*}$ ($\mathrm{FL}_-^{\varphi_*}$),
where each fixed point attracts a single orbit, constituting its
stable manifold (the two kinaton fixed points $\mathrm{K}^\pm_\pm$ are saddles). 
It is also helpful to note that
$\bar\varphi$ is increasing when $u>0$ and decreasing when $u<0$, which,
e.g., determines the flow directions along the boundaries $u = \pm\sqrt{2}$.

Third, the equations for $u$ and $v$ on the boundaries $\bar{\varphi} = \pm\pi$ are the same
as those for a constant potential and are easily solved. Furthermore, the straight lines
$\mathrm{FL}_0^\pm \rightarrow \mathrm{dS}^\pm$, connecting the boundary
$\mathrm{FL}_0^\pm$ fixed points at $(\bar{\varphi},u,v) = (\pm\pi,0,0)$
with the boundary de Sitter fixed points $\mathrm{dS}^\pm$ at $(\bar{\varphi},u,v) = (\pm\pi,0,1/\sqrt{3})$,
describe the $\Lambda$CDM model. There is also an interior $\Lambda$CDM orbit
$\mathrm{FL}_0^0\rightarrow \mathrm{dS}^0$ given by $\bar{\varphi} = 0$, $u = 0$,
connecting a fixed point on the line $\mathrm{FL}_0^{\varphi_*}$ at $(\bar{\varphi},u,v) = (0,0,0)$
with the de Sitter fixed point $\mathrm{dS}^0$ at $(\bar{\varphi},u,v) = (0,0,1/\sqrt{3})$.
Recall that for the $\Lambda$CDM solution (where now $\Omega_\varphi$ replaces $\Omega_\Lambda$)
\begin{equation}
\Omega_{\varphi} = \frac{\Omega_{\varphi 0}}{\Omega_{\mathrm{m}0}e^{-3N} + \Omega_{\varphi 0}},\qquad
\Omega_{\mathrm{m}0} + \Omega_{\varphi 0} = 1,
\end{equation}
which yields $v$ since $v = \sqrt{\Omega_\varphi/3}$.

\begin{figure}[ht!]
	\begin{center}
		\subfigure[The invariant boundary $\Omega_\mathrm{m}=0$ with $\bar{m}=1$.]{\label{PSFB1i}
			\includegraphics[width=0.30\textwidth]{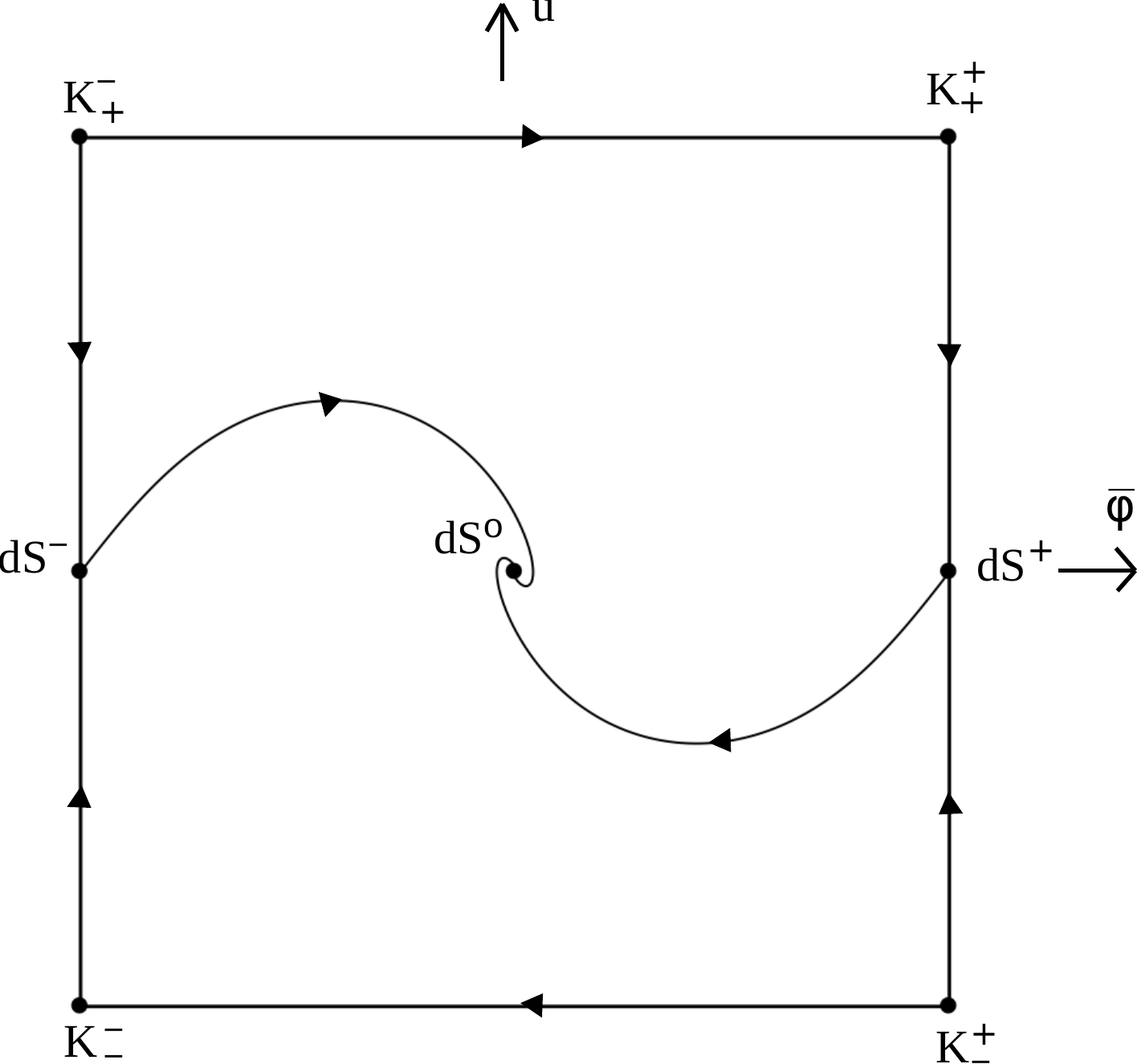}}
		\hspace{2.0cm}
		\subfigure[The invariant boundary $\Omega_\mathrm{m}=0$ with $\bar{m}=10$.]{\label{PSFB2i}
			\includegraphics[width=0.30\textwidth]{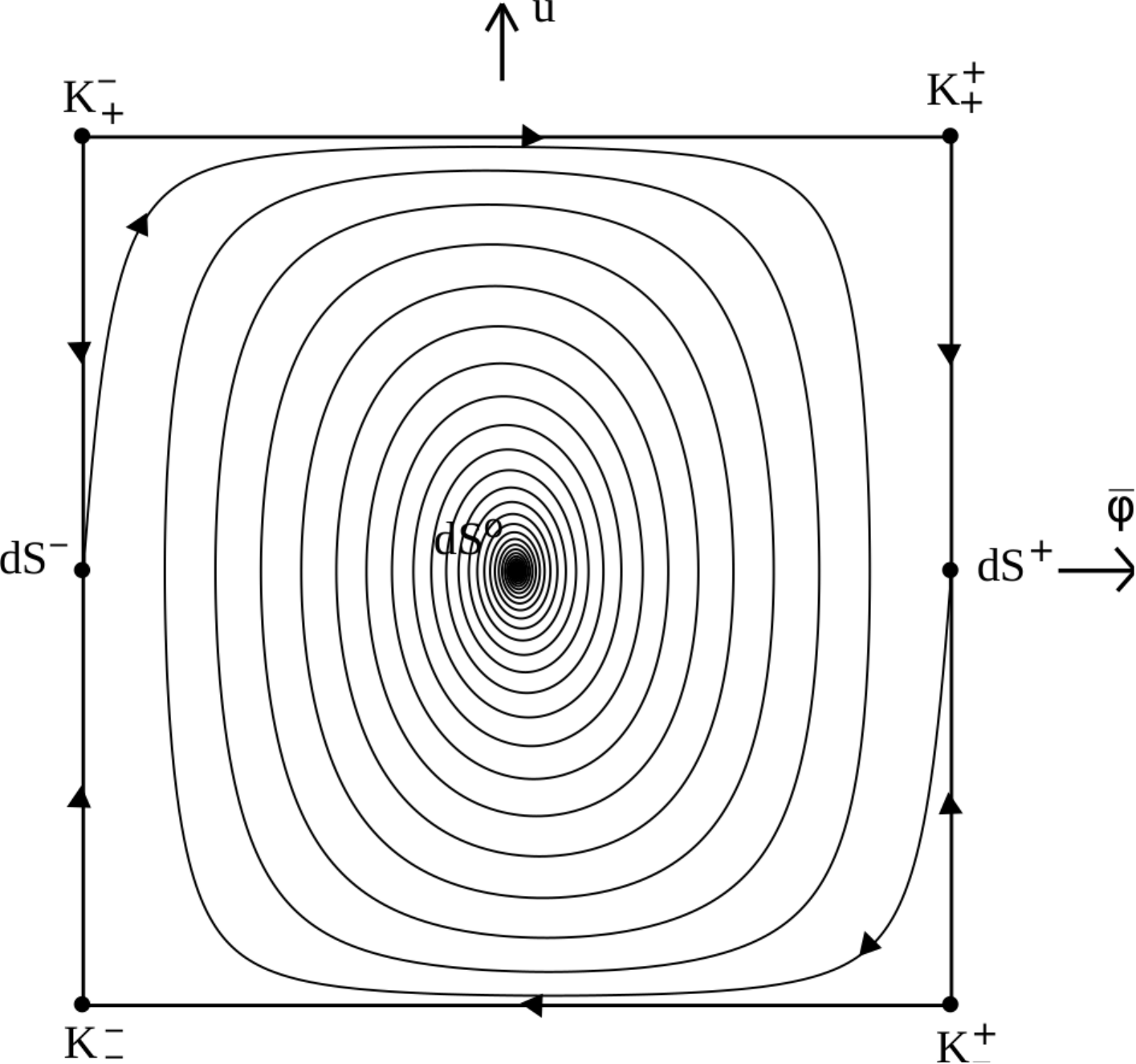}}
    \vspace{-0.5cm}
	\end{center}
	\caption{Inflationary attractor (separatrix) orbits $\mathrm{dS}^\pm \rightarrow \mathrm{dS}^0$ on the $v= 1/\sqrt{3}$
    ($\Omega_\varphi = 1$) boundary for $V(\varphi) = \Lambda(1  + \frac12 \bar{m}^2\,\varphi^2)$,
    where $\bar{m} = \frac{m}{\sqrt{2\Lambda}}$.
    }
	\label{PotV2Pos}
\end{figure}

The de Sitter fixed point $\mathrm{dS}^0$ is a hyperbolic sink where all interior orbits end,
including those on the scalar field boundary $v=1/\sqrt{3}$ ($\Omega_\mathrm{m} = 0$,
$\Omega_\varphi = 1$). Thus, $\mathrm{dS}^0$ is the future attractor of the present models;
see~\cite{waiell97} for a rigorous definition of the concept attractor.
The orbits $\mathrm{dS}^\pm \rightarrow \mathrm{dS}^0$ on the $v= 1/\sqrt{3}$
($\Omega_\varphi = 1$) boundary are the two (due to the discrete symmetry) equivalent
\emph{inflationary attractor solutions}, see Figure~\ref{PotV2Pos}. These orbits are the unstable
\emph{center manifold} of the fixed points $\mathrm{dS}^\pm$, while these fixed points
are sinks \emph{on} the $\bar{\varphi} = \pm \pi$ boundaries. This latter feature pushes
nearby orbits toward the inflationary attractor solutions where the zero eigenvalue strengthens
its attracting nature and also enables solutions to stay in a long inflationary quasi-de Sitter
epoch in the vicinity of $\mathrm{dS}^\pm$; cf. the discussion and figures in~\cite{feletal02}.

\begin{figure}[ht!]
	\begin{center}
		\subfigure[$\bar{m}=1$.]{\label{3DSS1i}
			\includegraphics[width=0.4\textwidth]{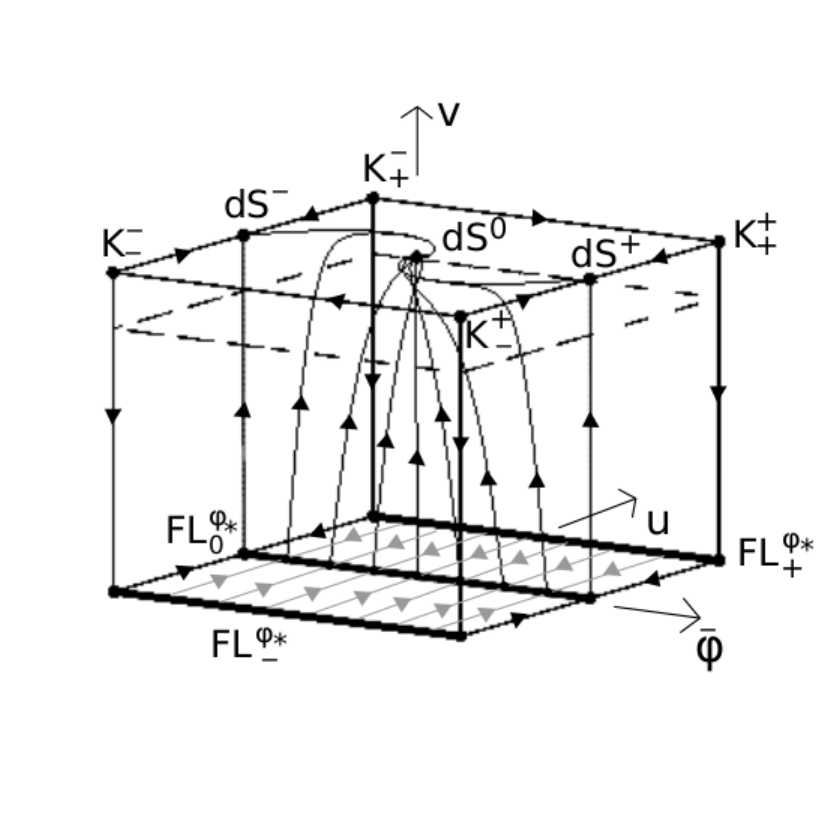}}
		\hspace{2.0cm}
		\subfigure[$\bar{m}=10$.]{\label{3DSS2i}
			\includegraphics[width=0.4\textwidth]{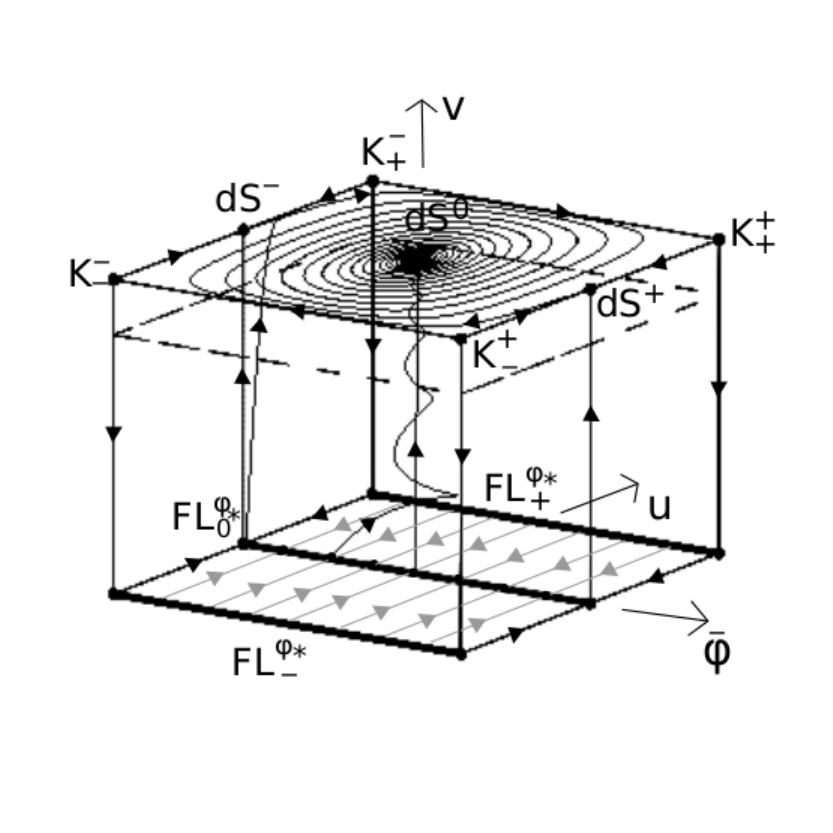}}
		\vspace{-0.5cm}
	\end{center}
	\caption{State space $(\bar{\varphi},u,v)$ depiction of thawing quintessence orbits for the potential
$V(\varphi) = \Lambda(1 + \bar{m}^2\,\varphi^2)$, where $\bar{m} = \frac{m}{\sqrt{2\Lambda}}$,
and orbits with frozen $\varphi =\varphi_*$ on the $v=0$ boundary, responsible for the thawing quintessence
mechanism in this state space representation.}
	\label{PotV2Posthaw}
\end{figure}

Of particular interest are the \emph{thawing quintessence attractor solutions}.
These are given by the one-parameter set of orbits that originates from $\mathrm{FL}_0^{\varphi_*}$,
one from each fixed point. These orbits form a two-dimensional surface of thawing quintessence
`attractor' orbits, illustrated in Figure~\ref{PotV2Posthaw}.
An observationally required long matter-dominated epoch requires
an open set of solutions to be extremely close to the $v=0$ boundary
where orbits are described by frozen values of $\bar{\varphi} = \bar{\varphi}_*$ ending at
$\mathrm{FL}_0^{\varphi_*}$, forming its stable manifold. Nearby orbits shadow these $v=0$ boundary
orbits and are thereby pushed toward the two-dimensional surface of thawing quintessence attractor orbits,
i.e. the unstable manifold of $\mathrm{FL}_0^{\varphi_*}$, which they subsequently
shadow. The thawing quintessence attractor orbits thereby describe the thawing quintessence epoch
for this open set of shadowing orbits. Finally, note that the inflationary attractor orbits, 
$\mathrm{dS}^\pm \rightarrow \mathrm{dS}^0$, form the upper $v= 1/\sqrt{3}$ boundary of the
surface of thawing quintessence attractor orbits, see Figure~\ref{PotV2Posthaw}.
%

%
%

\section{Models with $V=\Lambda - \frac12 m^2\varphi^2$\label{sec:Lm}}

Notably some of the physical properties of some of the solutions of the hilltop 
potential $V = \Lambda - \frac12 m^2\varphi^2$ were discussed in~\cite{feletal02}
and~\cite{Alam2003}, but here we will give a global dynamical systems account of 
the solution space of these models.

\subsection{Dynamical system derivation\label{subsec:dynsysderLm}}

To deal with these models we first note that~\eqref{Gauss} for the present potential can be written as
\begin{equation}
3H^2 + \frac12\,m^2\varphi^2 = \frac12\dot{\varphi}^2 + \Lambda + \rho_\mathrm{m}.
\end{equation}
Next we follow the idea introduced in~\cite{uggvon90} by defining the following `dominant' quantity
\begin{equation}
D = \sqrt{3H^2 + \frac12\,m^2\varphi^2} = \sqrt{\frac12\dot{\varphi}^2 + \Lambda + \rho_\mathrm{m}},
\end{equation}
which we use to define the following dimensionless bounded variables:
\begin{equation}\label{vardef}
\bar{s} = \frac{\sqrt{\Lambda}}{D},\quad \bar{H} = \frac{\sqrt{3}H}{D},\quad
\bar{\varphi} = \frac{m\varphi}{\sqrt{2}D},\quad \bar{\Sigma}_\varphi = \frac{\dot{\varphi}}{\sqrt{2}D},\quad
\bar{\Omega}_\mathrm{m} = \frac{\rho_\mathrm{m}}{D^2},
\end{equation}
where $\bar{s}>0$, $\bar{\Omega}_\mathrm{m} >0$.

It follows from the definitions in~\eqref{vardef} and the constraint~\eqref{Gauss} that
the 5D state vector $(\bar{s}, \bar{H}, \bar{\varphi}, \bar{\Sigma}_\varphi, \bar{\Omega}_\mathrm{m})$
satisfies the two constraints
\begin{subequations}\label{firstconstr}
\begin{align}
1 &= \bar{H}^2 + \bar{\varphi}^2,\label{compact}\\
1 &= \bar{\Sigma}_\varphi^2 + \bar{s}^2 + \bar{\Omega}_\mathrm{m},\label{Gausscompact}
\end{align}
\end{subequations}
while $(\bar{s}, \bar{H}, \bar{\varphi}, \bar{\Sigma}_\varphi, \bar{\Omega}_\mathrm{m})$
yield the following evolutions equations, as follows from the
definitions~\eqref{vardef} and the equations in~\eqref{Mainsysdim},
\begin{subequations}\label{firstdynsys}
\begin{align}
\frac{d\bar{s}}{d\bar{\tau}} &= Q\bar{s},\\
\frac{d\bar{H}}{d\bar{\tau}} &= Q\bar{H} - \left(2\bar{\Sigma}_\varphi^2 + \bar{\Omega}_\mathrm{m}\right),\\
\frac{d\bar{\varphi}}{d\bar{\tau}} &= Q\bar{\varphi} + \tilde{m}\,\bar{s}\,\bar{\Sigma}_\varphi,\\
\frac{d\bar{\Sigma}_\varphi}{d\bar{\tau}} &= (Q - 2\bar{H})\bar{\Sigma}_\varphi + \tilde{m}\,\bar{s}\,\bar{\varphi}, \\
\frac{d\bar{\Omega}_\mathrm{m}}{d\bar{\tau}} &= 2(Q - \bar{H})\bar{\Omega}_\mathrm{m}.
\end{align}
\end{subequations}
Here we have defined
\begin{equation}
Q = \left(2\bar{\Sigma}_\varphi^2 + \bar{\Omega}_\mathrm{m}\right)\bar{H} - \tilde{m}\,\bar{s}\,\bar{\varphi}\,\bar{\Sigma}_\varphi,
\end{equation}
\begin{equation}
\tilde{m} = \frac{2m}{\sqrt{3\Lambda}} = 2\sqrt{\frac{2}{3}}\,\bar{m},
\end{equation}
and introduced a new dimensionless time variable $\bar{\tau}$ defined by
\begin{equation}
\frac{d\bar{\tau}}{dt} = \frac{\sqrt{3}}{2}\,D,
\end{equation}
from which it follows that $\bar{\tau}$ is related to the $e$-fold time $N$
according to
\begin{equation}
\frac{dN}{d\bar{\tau}} = \frac23\,\bar{H}.
\end{equation}

The constrained dynamical system~\eqref{firstconstr}, \eqref{firstdynsys} admits the following
discrete symmetries
\begin{subequations}\label{2discr}
\begin{align}
(\bar{\varphi},\bar{\Sigma}_\varphi) &\mapsto -(\bar{\varphi},\bar{\Sigma}_\varphi),\label{2discr1}\\
(\bar{\tau},\bar{H},\bar{\varphi}) &\mapsto -(\bar{\tau},\bar{H},\bar{\varphi}),\label{2discr2}
\end{align}
\end{subequations}
where combining these two discrete symmetries yield the discrete symmetry
$(\bar{\tau},\bar{H},\bar{\Sigma}_\varphi) \mapsto -(\bar{\tau},\bar{H},\bar{\Sigma}_\varphi)$,
which illustrates that the present state space admits time inversion symmetry
(this is also the case for positive scalar field potentials, but there we pick out
the expanding component; this, however, is not possible for
potentials that can change sign since they only yield a single connected state space),
whereas~\eqref{2discr1} is a consequence of the discrete symmetry
$\bar{\varphi} \mapsto -\bar{\varphi}$ of the potential.

It is  of interest to consider the following quantities in the new dimensionless bounded variables:
\begin{subequations}\label{relations1}
\begin{alignat}{3}
\varphi &= 2\sqrt{\frac{2}{3}}\,\tilde{m}^{-1}\left(\frac{\bar{\varphi}}{\bar{s}}\right),&\qquad
V &= \Lambda\left[1 - \left(\frac{\bar{\varphi}}{\bar{s}}\right)^2\right],\\
H &= \sqrt{\frac{\Lambda}{3}}\left(\frac{\bar{H}}{\bar{s}}\right),&\qquad
\Omega_\mathrm{m} &= \frac{\bar{\Omega}_\mathrm{m}}{\bar{H}^2}, \label{H_Ommrel}\\
w_\varphi &= \frac{\bar{\Sigma}_\varphi^2 + (\bar{\varphi}^2 - \bar{s}^2)}{\bar{\Sigma}_\varphi^2 - (\bar{\varphi}^2 - \bar{s}^2)},&\quad
q &= -1 + \frac32\left(\frac{2\bar{\Sigma}_\varphi^2 +\bar{\Omega}_\mathrm{m}}{\bar{\varphi}^2}\right).\label{wvarphi_q}
\end{alignat}
\end{subequations}
%

%
%

Next we solve the constraint~\eqref{compact} as follows:
\begin{equation}
\bar{H} = \cos\theta,\qquad \bar{\varphi} =\sin\theta,
\end{equation}
while the constraint~\eqref{Gausscompact} is used to globally solve
for $\bar{\Omega}_\mathrm{m}$, i.e.,
\begin{equation}
\bar{\Omega}_\mathrm{m} = 1 - \bar{\Sigma}_\varphi^2 - \bar{s}^2,
\end{equation}
which leads to the following unconstrained three-dimensional regular dynamical system
\begin{subequations}\label{dynsys}
\begin{align}
\frac{d\bar{s}}{d\bar{\tau}} &= Q\bar{s},\\
\frac{d\bar{\Sigma}_\varphi}{d\bar{\tau}} &= (Q - 2\cos\theta)\bar{\Sigma}_\varphi + \tilde{m}\bar{s}\sin\theta,\\
\frac{d\theta}{d\bar{\tau}} &= (1 + \bar{\Sigma}_\varphi^2 - \bar{s}^2)\sin\theta + \tilde{m}\bar{s}\bar{\Sigma}_\varphi\cos\theta,
\end{align}
\end{subequations}
where
\begin{equation}
Q = (1 + \bar{\Sigma}_\varphi^2 - \bar{s}^2)\cos\theta -\tilde{m}\bar{s}\bar{\Sigma}_\varphi\sin\theta.
\end{equation}
It is also of interest to consider the now auxiliary equation
\begin{equation}
\frac{d\bar{\Omega}_\mathrm{m}}{d\bar{\tau}} = 2(Q - \cos\theta)\bar{\Omega}_\mathrm{m},
\end{equation}
which shows that $\bar{\Omega}_\mathrm{m} = 1 - \bar{\Sigma}_\varphi^2 - \bar{s}^2 = 0$ is an invariant
boundary subset, as is $\bar{s}=0$.
Together these boundary subsets form the boundary of the state space for the
state vector $(\bar{s},\bar{\Sigma}_\varphi,\theta)$. Thanks to the regularity of the dynamical system we now include
these boundaries yielding a compact state space for $(\bar{s},\bar{\Sigma}_\varphi,\theta)$.
The state space is thereby given by a torus sliced in half, where the bottom of the sliced torus is given
by the invariant boundary set $\bar{s}=0$ whereas its upper boundary is the invariant set
$\bar{\Omega}_\mathrm{m} = 1 - \bar{\Sigma}_\varphi^2 - \bar{s}^2 = 0$.
The discrete symmetries in~\eqref{2discr} take the following form in the system~\eqref{dynsys}:
\begin{subequations}\label{discretemaster}
\begin{align}
(\bar{\Sigma}_\varphi,\theta) &\mapsto -(\bar{\Sigma}_\varphi,\theta),\label{D1}\\
(\bar{\tau},\theta) &\mapsto (-\bar{\tau}, \theta - (2n+1)\pi),\qquad n\in\mathbb{Z}\label{D2}
\end{align}
\end{subequations}
where $\theta \mapsto \theta - (2n+1)\pi$ results in
$(\cos\theta,\sin\theta) \mapsto -(\cos\theta,\sin\theta)$.
Due to the two discrete symmetries, one can always relate one
interior state space orbit to three others, except for when
$\bar{\Sigma}_\varphi = 0$, which yields
the two `interior' $\Lambda$CDM orbits with $\bar{\Sigma}_\varphi = 0$, 
$\sin\theta = 0$ and the two boundary orbits with
$\bar{\Sigma}_\varphi = 0$, $\bar{s}=0$, discussed below.

\subsection{Dynamical systems analysis\label{subec:dynsysanalysismp}}

We first note that
\begin{equation}\label{Mdef}
M = \sqrt{\frac{3}{\Lambda}}\,H = \frac{\bar{H}}{\bar{s}} = \frac{\cos\theta}{\bar{s}}
\end{equation}
is a strictly monotonically decreasing function in the interior state space, i.e,
when $\bar{s}>0$ and $\bar{\Omega}_\mathrm{m}>0$, since~\eqref{firstdynsys} leads to
\begin{equation}\label{Meq}
\frac{dM}{d\bar{\tau}} = -\frac{2\bar{\Sigma}_\varphi^2 + \bar{\Omega}_\mathrm{m}}{\bar{s}}
= -\frac{1 + \bar{\Sigma}_\varphi^2 - \bar{s}^2}{\bar{s}} < 0,
\end{equation}
and hence there are no interior state space fixed points or recurring orbits,
which brings the boundary subsets of the state space into focus.

\subsubsection*{The invariant $\bar{s} = 0$ boundary set}

The equations on the $\bar{s} = 0$ boundary are obtained from~\eqref{dynsys} by
setting $\bar{s}=0$, which results in
\begin{subequations}\label{dynsyss0}
\begin{align}
\frac{d\bar{\Sigma}_\varphi}{d\bar{\tau}} &= -(1 - \bar{\Sigma}_\varphi^2)\bar{\Sigma}_\varphi\cos\theta,\\
\frac{d\theta}{d\bar{\tau}} &= (1 + \bar{\Sigma}_\varphi^2)\sin\theta,
\end{align}
\end{subequations}
which thereby yields a system of equations that are independent of $\tilde{m}$.
The state space is a circular strip with $\bar{\Sigma}_\varphi = -1$ as its invariant inner
circle boundary and $\bar{\Sigma}_\varphi = +1$ as the invariant outer circle boundary. Due to the discrete
symmetries in~\eqref{discretemaster}, the state space is divided into four disjoint invariant regions with
the invariant boundary subsets and the invariant inner subsets $\bar{\Sigma}_\varphi = 0$;
$\theta = 2n\pi$ and $\theta = (2n+1)\pi$, $n\in\mathbb{Z}$, i.e., $\sin\theta=0$, $\cos\theta = \pm1$,
as their boundaries.

The intersections of these one-dimensional invariant sets form 6 fixed points.
There are four scalar field kinetic energy dominated \emph{kinaton fixed points} given by
\begin{equation}\label{K}
\mathrm{K}^\pm_\epsilon = \left\{ (\bar{s}, \bar{\Sigma}_\varphi, \cos\theta)
= (0,\epsilon,\pm 1) \right\},
\end{equation}
where $\epsilon = 1$ or $\epsilon = -1$.
The superscript in $\mathrm{K}^\pm_\epsilon$ denotes if $\cos\theta = \bar{H}$
is equal to $1$ or $-1$, a nomenclature that we will use throughout, while
the value of the subscript $\epsilon$ denotes if $\bar{\Sigma}_\varphi = -1$ or if
$\bar{\Sigma}_\varphi = +1$. The discrete symmetry~\eqref{D1} results in that
the fixed points with the same superscript have the same eigenvalues irrespective
of the sign of $\epsilon$ while the discrete symmetry~\eqref{D2} yields
eigenvalues with opposite signs for fixed points with the same value of
$\epsilon$ but with an opposite sign of $\bar{H}$. On the boundary
$\bar{s}=0$ the fixed points $\mathrm{K}^+_\pm$ are hyperbolic sources while
$\mathrm{K}^-_\pm$ are hyperbolic sinks.

\begin{figure}[ht!]
	\begin{center}
			\includegraphics[trim={0cm 0cm 0cm 0cm},width=0.3\textwidth]{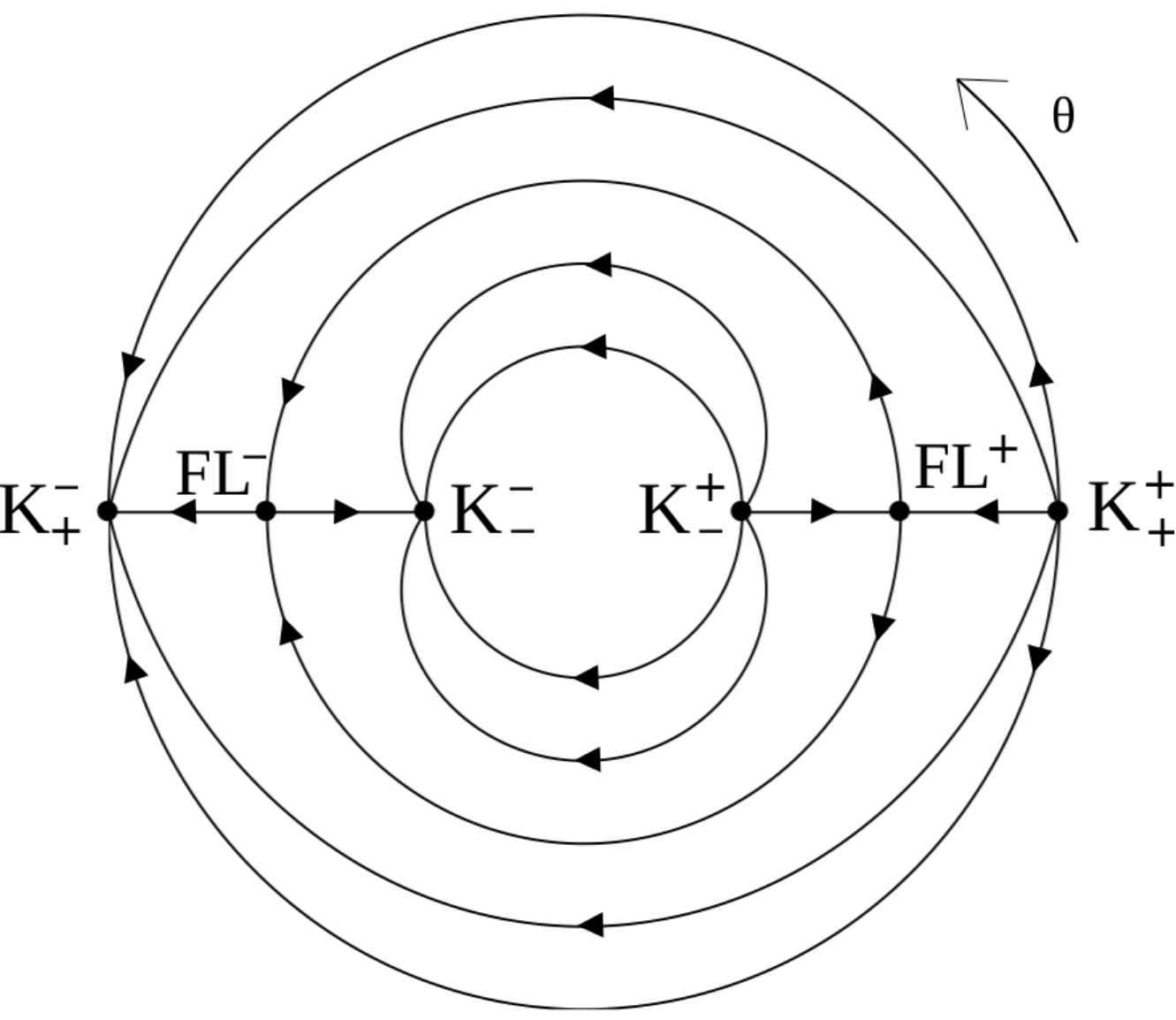}
				\hspace{2.0cm}
			\includegraphics[trim={0cm 0cm 0cm 0cm},width=0.3\textwidth]{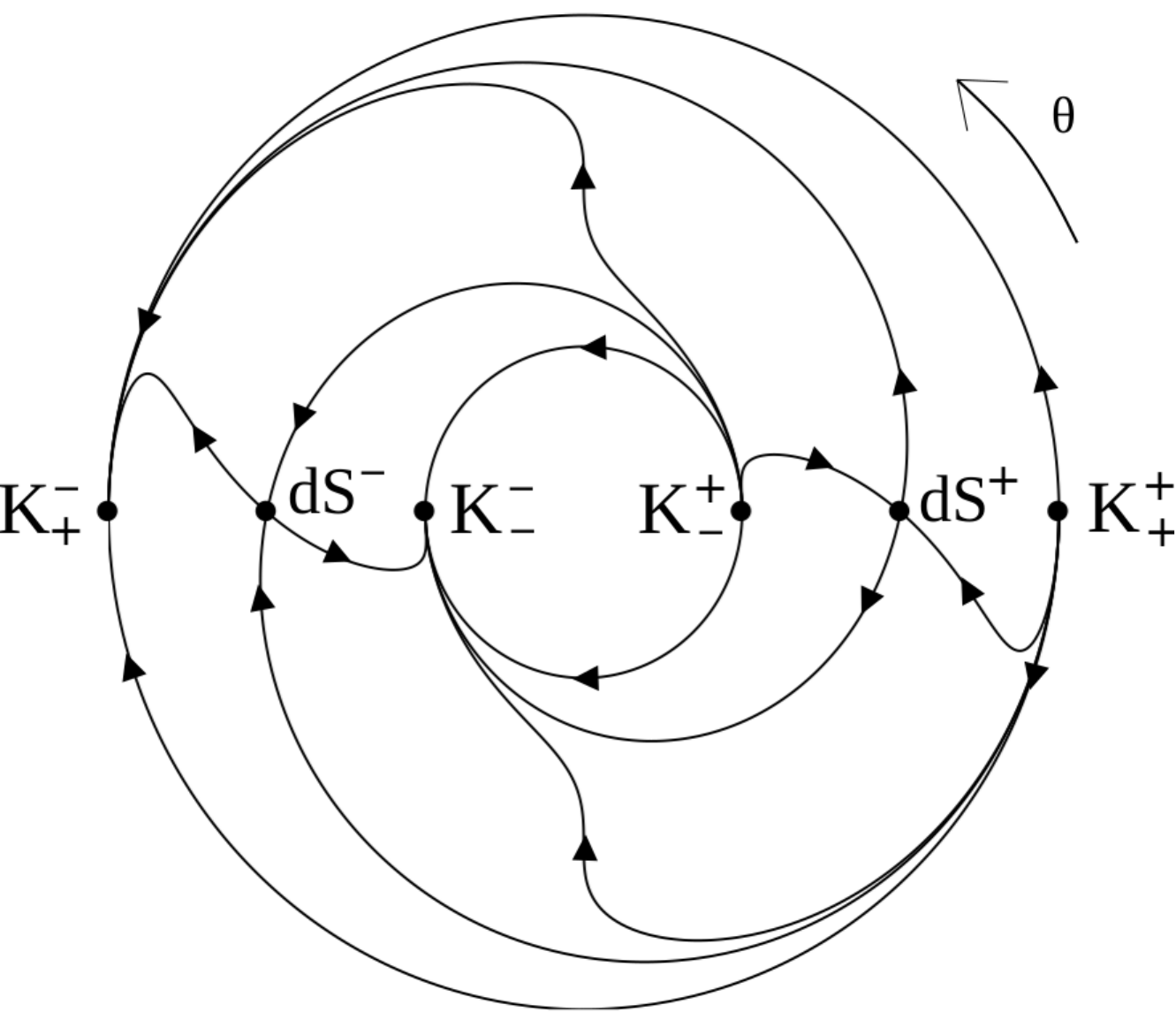}
    \vspace{-0.5cm}
	\end{center}
	\caption{Orbit structure on the $\bar{s}=0$ (figure to the left) and $\bar{\Omega}_\mathrm{m}=0$
    (figure to the right) boundaries, where in the latter case the orbits have been projected onto the
    plane $\bar{s}=0$. }
	\label{boundaries}
\end{figure}

In addition to the four kinaton fixed points there are
the two \emph{Friedman-Lema\^itre fixed points} given by
\begin{equation}\label{FL}
\mathrm{FL}^\pm = \left\{ (\bar{s}, \bar{\Sigma}_\varphi, \cos\theta)
= (0,0,\pm 1) \right\}.
\end{equation}
On the $\bar{s}=0$ boundary both these fixed points are hyperbolic saddles.
Finally, we note that the dynamical system~\eqref{dynsyss0} is integrable,
admitting the following conserved quantity
\begin{equation}
\bar{\Sigma}_\varphi\sin\theta = k(1 - \bar{\Sigma}_\varphi^2).
\end{equation}
Note the invariant $\bar{\Sigma}_\varphi=0$ $(k=0)$
subset for which $\bar{\Omega}_\mathrm{m}=1$ and $w_\varphi = -1$.
The orbit structure on $\bar{s}=0$ is given in Figure~\ref{boundaries}.
%

\subsubsection*{The invariant $\bar{\Omega}_\mathrm{m} = 0$ boundary set}

The equations on this subset are obtained by setting
$\bar{\Omega}_\mathrm{m} = 1 - \bar{\Sigma}_\varphi^2 - \bar{s}^2 = 0$ in~\eqref{dynsys}.
In this case $M = \bar{H}/\bar{s} = \cos\theta/\bar{s}$ yields
\begin{equation}
\frac{dM}{d\bar{\tau}} = -\frac{2\bar{\Sigma}_\varphi^2}{\bar{s}} = -\frac{2(1 - \bar{s}^2)}{\bar{s}};\qquad
\left. \frac{d^2M}{d\bar{\tau}^2}\right|_{\bar{s}=1} = 0,\qquad
\left. \frac{d^3M}{d\bar{\tau}^3}\right|_{\bar{s}=1} = - 4\tilde{m}^2\sin^2\theta.
\end{equation}
Thus, $M$ is monotonically decreasing everywhere on this subset except possibly at
$\bar{s}=1$, $\bar{\Sigma}_\varphi =0$; but, as follows from the last two of the above
equations, $M$ is also monotonically decreasing when
$\bar{s} = 1$, $\bar{\Sigma}_\varphi =0$,
\emph{except at} the two \emph{de Sitter fixed points} with $\sin\theta=0$,
\begin{equation}\label{K}
\mathrm{dS}^\pm = \left\{ (\bar{s}, \bar{\Sigma}_\varphi, \cos\theta)
= (1,0,\pm 1) \right\},
\end{equation}
where both are hyperbolic saddles on the boundary subset $\bar{\Omega}_\mathrm{m} = 0$;
for the orbit structure, see Figure~\ref{boundaries}.

\subsubsection*{The interior invariant $\Lambda\mathrm{CDM}^\pm$ sets}

In addition to the above boundary subsets there are also the interior expanding and contracting
$\Lambda\mathrm{CDM}^\pm$ subsets given by two orbits parameterized by $\bar{s}$ according to
\begin{subequations}
\begin{alignat}{5}
\Lambda\mathrm{CDM}^+:&\quad \bar{\Sigma}_\varphi &= 0,&\qquad \sin\theta &= 0,&\qquad\cos\theta &= 1,&\qquad \frac{d\bar{s}}{d\bar{\tau}} &= (1 - \bar{s}^2)\bar{s}, \\
\Lambda\mathrm{CDM}^-:&\quad \bar{\Sigma}_\varphi &= 0,&\qquad \sin\theta &= 0,&\qquad\cos\theta &= -1, &\qquad \frac{d\bar{s}}{d\bar{\tau}} &= -(1 - \bar{s}^2)\bar{s},
\end{alignat}
\end{subequations}
respectively, i.e., $\theta = 2n\pi$ ($\theta = (2n + 1)\pi$), $n\in\mathbb{Z}$, for the expanding $\Lambda\mathrm{CDM}^+$
(contracting $\Lambda\mathrm{CDM}^-$) subset. Note that $\bar{\Omega}_\mathrm{m} = 1 - \bar{s}^2$ and hence that
$\frac{d\bar{\Omega}_\mathrm{m}}{d\bar{\tau}} = \mp 2(1 - \bar{\Omega}_\mathrm{m})\bar{\Omega}_\mathrm{m}$
for these two subsets. Note further that the expanding $\Lambda\mathrm{CDM}^+$
(contracting $\Lambda\mathrm{CDM}^-$) subset is given by the heteroclinic orbit
$\mathrm{FL}^+ \rightarrow \mathrm{dS}^+$ ($\mathrm{dS}^- \rightarrow \mathrm{FL}^-$).

It follows from the monotonic function $M$ and the structure on the invariant boundary sets that
all interior orbits are heteroclinic orbits. We therefore now complete the dynamical systems
analysis by situating the stability analysis of the fixed points in the full three-dimensional
state space $(\bar{s},\bar{\Sigma}_\varphi,\theta)$. Note, however, that this analysis follows
from combining the results of the above subsets, in particular, use Figure~\ref{boundaries}
in combination with the $\Lambda\mathrm{CDM}^\pm$ heteroclinic orbits
$\mathrm{FL}^+ \rightarrow \mathrm{dS}^+$ and $\mathrm{dS}^- \rightarrow \mathrm{FL}^-$
to obtain an intuitive picture.

\subsubsection*{Fixed points and their stability}

All fixed points satisfy $\bar{\varphi} = \sin\theta=0$ and
$\bar{H} = \cos\theta = \pm 1$ and due to the discrete symmetry~\eqref{D2}
it follows that they always appear in pairs, one corresponding to
expansion and one to contraction, where the absolute values of the
eigenvalues are the same but where $\bar{\tau}\mapsto -\bar{\tau}$
in~\eqref{D2} leads to that they have opposite signs.
\begin{itemize}
\item[$\boxed{\mathrm{K}^\pm_\epsilon}$]
Due to the discrete symmetries~\eqref{discretemaster}
it suffices to calculate the eigenvalues of one of the four 
\emph{kinaton fixed points} since the eigenvalues for the other ones 
then follow trivially. The kinaton fixed points
$\mathrm{K}^+_\pm$ are hyperbolic sources while
$\mathrm{K}^-_\pm$ are hyperbolic sinks.
\item[$\boxed{\mathrm{dS}^\pm}$] The \emph{de Sitter fixed point}
$\mathrm{dS}^+$ is a hyperbolic saddle with an eigenvalue
$-2$ with an eigenvector along the expanding $\Lambda\mathrm{CDM}^+$ orbit
and two eigenvalues $\pm \sqrt{1 + \tilde{m}^2} - 1$ with eigenvectors
on the scalar field boundary $\bar{\Omega}_\mathrm{m} = 0$, forming
a saddle on this boundary, see Figure~\ref{boundaries}.
Thus, there is a one-parameter set of orbits ending at 
$\mathrm{dS}^+$.\footnote{This one-parameter set of orbits, 
as well as those mentioned below, has two branches related by a 
discrete symmetry; for brevity, however, we only refer to them 
as one-parameter sets of orbits.} 
As follows from~\eqref{D2}, the eigenvalues
of $\mathrm{dS}^-$ have the opposite signs and is thereby also a hyperbolic
saddle; thus there is just one negative eigenvalue with an eigenvector
on the $\bar{\Omega}_\mathrm{m}=0$ boundary, whereas a one-parameter set
of orbits originates from $\mathrm{dS}^-$.
%
\item[$\boxed{\mathrm{FL}^\pm}$] Among all the fixed points,
the \emph{Friedmann-Lema\^{i}tre fixed point} $\mathrm{FL}^+$
is the most interesting one from a quintessence perspective since it is from this
fixed point the one-parameter set of quintessence attractor orbits originate from.
Note further that the attractor mechanism of these orbits is given by the
stable manifold of $\mathrm{FL}^+$, given by the orbits
$\mathrm{K}_\pm^+ \rightarrow \mathrm{FL}^+$ with $\bar{s} = 0$,
$\theta=2n\pi$, $n\in\mathbb{Z}$, (see Figure~\ref{boundaries}), which pushes an 
open set of nearby orbits toward the unstable manifold of quintessence 
attractor orbits, which they thereafter shadow and thereby are approximated
by. Note further that a long matter-dominated epoch requires solutions to be very
close to the $\mathrm{FL}^+$ fixed point.

Finally, note that there is a one-parameter set of orbits ending at the
fixed point $\mathrm{FL}^-$ (its stable manifold), including the $\Lambda\mathrm{CDM}^-$ orbit,
although nearby orbits are pushed away from this subset of orbits by the unstable manifold,
given by the orbits $\mathrm{FL}^- \rightarrow \mathrm{K}_\pm^-$ with
$\bar{s} = 0$, $\theta=(2n + 1)\pi$, $n\in\mathbb{Z}$, toward the sinks $\mathrm{K}^-_\pm$.
\end{itemize}

Let us now take a closer look at the quintessence attractor subset of orbits
and let us without loss of generality set $\theta = 0$. Then a linear analysis of
$\mathrm{FL}^+$ yield the following linear solution
\begin{equation}\label{FLlinj1}
\bar{s} = \bar{s}_1e^{\bar{\tau}},\qquad \bar{\Sigma}_\varphi = \sigma_1e^{-\bar{\tau}},\qquad \theta = \theta_1e^{\bar{\tau}}.
\end{equation}
%
The quintessence attractor orbits are given by the unstable manifold of $\mathrm{FL}^+$,
i.e. by setting $\sigma_1=0$. Recall that matter-dominance requires orbits to be close to
$\mathrm{FL}^+$, where they are pushed toward the unstable manifold of quintessence attractor
orbits by the stable manifold associated with the eigenvalue $-1$, corresponding to the
orbits $\mathrm{K}_\pm^+ \rightarrow \mathrm{FL}^+$ with $\bar{s}=0$ and $\theta = 0$.

It is of interest to obtain a better approximation for the unstable manifold than
the linear approximation in~\eqref{FLlinj1} by series expanding in $\exp(\bar{\tau})$.
However, since the variables in the constrained dynamical system~\eqref{firstconstr},
\eqref{firstdynsys} are more closely connected to the physical quantities we are
interested in and since it is more natural to use approximations for $\bar{H}$
and $\bar{\varphi}$ rather than $\theta$ we make a series expansion in
$\exp(\bar{\tau})$ for this system; when truncated at $\exp(3\bar{\tau})$ this yields
\begin{subequations}\label{FLlinj2}
\begin{align}
\bar{s} &\approx \bar{s}_1e^{\bar{\tau}}\left[1 -\frac{1}{4}(2\bar{s}^2_1+\bar{\varphi}^2_1)e^{2\bar{\tau}}\right],\\
\bar{H} &\approx 1 - \frac12\bar{\varphi}_1^2e^{2\bar{\tau}},\\ 
\bar{\varphi} &\approx \bar{\varphi}_1e^{\bar{\tau}}\left[1
 -\left(\frac{\bar{\varphi}^2_1}{4}+\frac{\bar{s}^2_1}{6}(3-\tilde{m}^2)\right)e^{2\bar{\tau}}\right],\\
\bar{\Sigma}_\varphi &\approx \frac13\tilde{m}\bar{s}_1\bar{\varphi}_1e^{2\bar{\tau}},\\
\bar{\Omega}_\mathrm{m} &\approx 1 - \bar{s}_1^2e^{2\bar{\tau}},
\end{align}
\end{subequations}
which leads to
\begin{subequations}
\begin{align}
\varphi &\approx \varphi_*\left(1 + \frac16\tilde{m}^2\bar{s}_1^2\,e^{2\bar{\tau}}\right),\\
w_\varphi &\approx -1 + \frac{2\tilde{m}^2\bar{s}_1^2\bar{\varphi}_1^2}{9(\bar{s}_1^2 - \bar{\varphi}_1^2)}\,e^{2\bar{\tau}},\\
\Omega_\mathrm{m} &\approx 1 - (\bar{s}^2_1 -\bar{\varphi}^2_1)\,e^{2\bar{\tau}},\\
V &\approx \Lambda\left[1 - \left(\frac{\bar{\varphi}_1}{\bar{s}_1}\right)^2\left(1 + \frac13\tilde{m}^2\bar{s}_1^2\,e^{2\bar{\tau}}\right)\right],
\end{align}
\end{subequations}
where $\varphi_* \equiv 2\sqrt{\frac23}\bar{\varphi}_1/(\tilde{m}\bar{s}_1)$ and
hence $\Lambda(\bar{\varphi}_1/\bar{s}_1)^2 = \frac12m^2\varphi_*^2$.

%
%
%

\begin{figure}[ht!]
	\begin{center}
			\includegraphics[trim={0cm 0cm 0cm 0cm},width=0.45\textwidth]{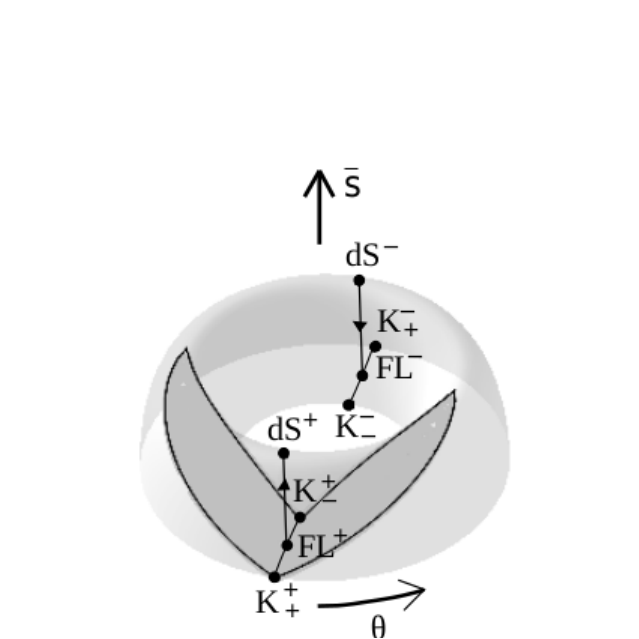}
	\end{center} 	
	\caption{The $V(\varphi)=0$ surface `originating' from $\theta=0$; in addition, there is a corresponding
$V(\varphi)=0$ surface, not depicted in order to avoid clutter, at $\theta=\pi$, since
$V=0 \Rightarrow \bar{s} = |\bar{\varphi}| = |\sin\theta|$.}
	\label{fig:T2V0}
\end{figure}

The potential is thereby past asymptotically positive when $\bar{s}_1^2 > \bar{\varphi}_1^2$,
see Figure~\ref{fig:T2V0} for the depiction of $V(\varphi)=0$ in the state space.
Moreover, $\Omega_\varphi$ and $1 + w_\varphi$ are positive since
$\Omega_\varphi \approx (\bar{s}^2_1 -\bar{\varphi}^2_1)\,e^{2\bar{\tau}}$ and
$1 + w_\varphi \approx \frac{2\tilde{m}^2\bar{s}_1^2\bar{\varphi}_1^2}{9(\bar{s}_1^2 - \bar{\varphi}_1^2)}\,e^{2\bar{\tau}}$.
The condition $\bar{s}_1^2 > \bar{\varphi}_1^2$ thereby leads to a one-parameter
set of \emph{thawing quintessence attractor solutions} with $w_\varphi$
initially increasing from the past asymptotic value $w_\varphi = -1$ and where $\Omega_\mathrm{m}$
is initially decreasing from its past asymptotic matter-dominated value $\Omega_\mathrm{m} = 1$,
where $\bar{\varphi}_1 = 0$ ($\varphi_* = 0$) corresponds to the expanding $\Lambda\mathrm{CDM}^+$ orbit.
However, the potential is past asymptotically negative when $\bar{s}_1^2 < \bar{\varphi}_1^2$,
which leads to that $\Omega_\varphi$ and $1 + w_\varphi$ decrease to negative values from their
past asymptotic zero values, i.e., this leads to a one-parameter set of
\emph{freezing quintessence attractor solutions} early
during their evolution.

Note, however, the following:
Apart from the $\Lambda\mathrm{CDM}^+$ orbit and the one-parameter set of orbits ending
with a big crunch at $\mathrm{FL}^-$, which includes the $\Lambda\mathrm{CDM}^-$ orbit, all orbits,
including the freezing quintessence attractor solutions, end at
one of the fixed points $\mathrm{K}^-_\pm$ where $w_\varphi = 1$. Furthermore, it follows
from the monotonic function $M$ and the local fixed point analysis and the orbit structure
on the $\bar{s} = 0$ and $\bar{\Omega}_\mathrm{m}=0$ boundaries that all initially expanding
solutions ($H>0$) originate from $\mathrm{K}^+_\pm$ (the generic case), $\mathrm{FL}^+$
(the one-parameter set of quintessence attractor solutions) and $\mathrm{dS}^+$
(the orbits $\mathrm{dS}^+ \rightarrow \mathrm{K}^-_\pm$ on the $\bar{\Omega}_\mathrm{m}=0$
boundary, see Figure~\ref{boundaries}).
These orbits end at $\mathrm{K}^-_\pm$ (the generic
case) and at $\mathrm{FL}^-$ (a one-parameter set, including the $\mathrm{FL}^+ \rightarrow \mathrm{FL}^-$
orbits on the $\bar{s} = \bar{\Sigma}_\varphi = 0$ subset, see Figure~\ref{boundaries})
and at $\mathrm{dS}^-$ (the orbits $\mathrm{K}^+_\pm\rightarrow \mathrm{dS}^-$ 
on the $\bar{\Omega}_\mathrm{m}=0$ boundary, see Figure~\ref{boundaries}).

\begin{figure}[ht!]
	\begin{center}
		\includegraphics[trim={0cm 0cm 0cm 0cm},width=0.45\textwidth]{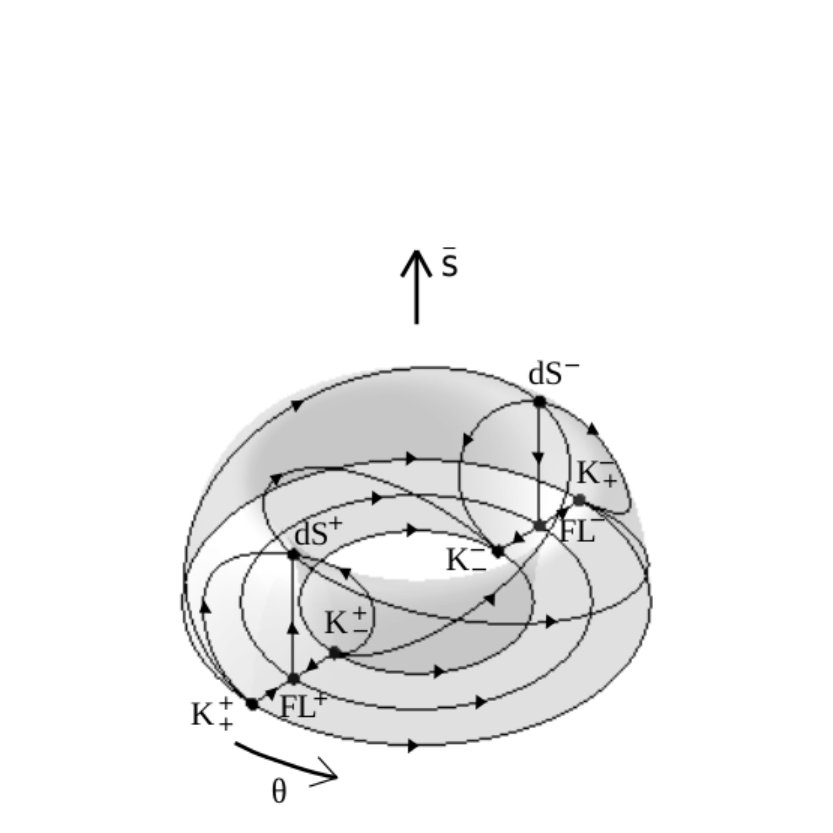}
	\end{center} 	
	\caption{Examples of orbits. The expanding and contracting $\Lambda\mathrm{CDM}^\pm$ orbits are given by
$\mathrm{FL}^+\rightarrow\mathrm{dS}^+$ and $\mathrm{dS}^-\rightarrow\mathrm{FL}^-$, respectively.
Viable thawing quintessence models more or less shadow one of the following two discrete
symmetry related heteroclinic orbit sequences
$\mathrm{FL}^+\rightarrow\mathrm{dS}^+\rightarrow\mathrm{K}_+^-$,
$\mathrm{FL}^+\rightarrow\mathrm{dS}^+\rightarrow\mathrm{K}_-^-$; they thereby end at a contracting
kinaton state with $\bar{H}=-1$ ($H\rightarrow - \infty$) and $w_\varphi = 1$.}
	\label{fig:T2}
\end{figure}

Thus all interior orbits ($\bar{s}>0$, $\bar{\Omega}_\mathrm{m}>0$), 
apart from the $\Lambda\mathrm{CDM}^+$ orbit, reach a point of maximum
expansion (where $H=0$ and thus $\bar{H}=\cos\theta = 0$, i.e., $\theta = \pi/2 + n\pi$, $n\in\mathbb{Z}$)
and then contract to a big crunch singularity.
%
As will be shown below, observationally viable thawing quintessence models more or less
shadow one of the following two discrete symmetry related heteroclinic orbit sequences
$\mathrm{FL}^+\rightarrow\mathrm{dS}^+\rightarrow\mathrm{K}_+^-$,
$\mathrm{FL}^+\rightarrow\mathrm{dS}^+\rightarrow\mathrm{K}_-^-$; they thereby end at a contracting
kinaton state with $\bar{H}=-1$ ($H\rightarrow - \infty$) and $w_\varphi = 1$.
Examples of orbits are given in Figure~\ref{fig:T2}. Compare these detailed results with those
in~\cite{feletal02}.
%

\subsection{The quintessence attractor subset\label{subsec:quintsubset}}

\begin{figure}[ht!]
	\begin{center}
		\includegraphics[trim={0cm 0cm 0cm 0cm},width=0.35\textwidth]{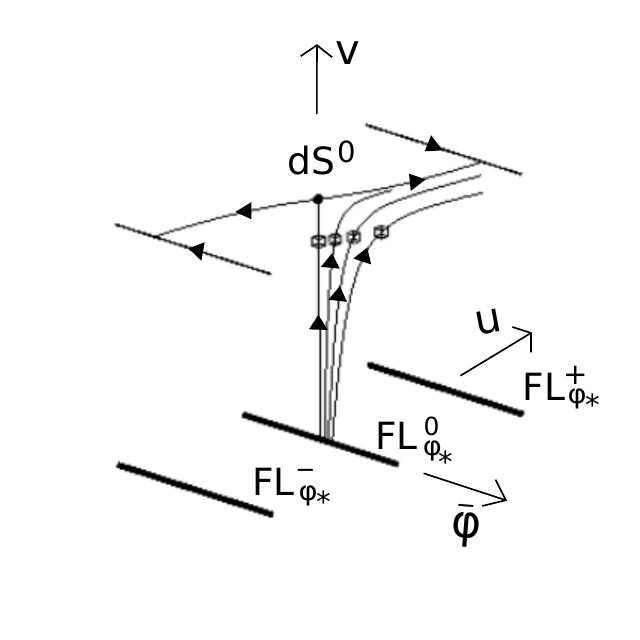}
		\includegraphics[trim={0cm 0cm 0cm 0cm},width=0.45\textwidth]{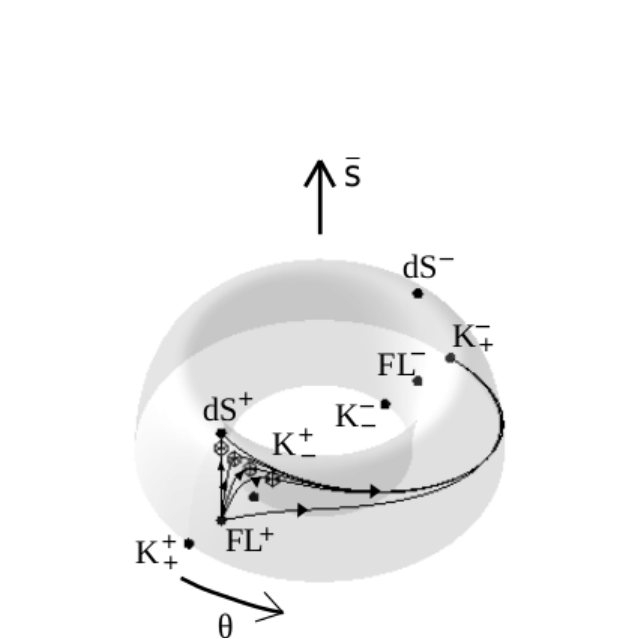}
	\end{center} 	
	\caption{Thawing quintessence attractor solutions (and the $\Lambda\mathrm{CDM}^+$ solution)
in the $(\bar{\varphi},u,v)$ state space and the corresponding ones
in the $(\bar{s},\bar{\Sigma}_\varphi,\theta)$ state space. The ring denotes when $\Omega_\varphi = 0.68$.}
	\label{fig:thaw2statespace}
\end{figure}

As shown above, the quintessence attractor subset is the one-parameter set of orbits forming the
unstable manifold of $\mathrm{FL}^+$. This subset is in turn divided into two parts: the thawing and
freezing orbits, where only the thawing quintessence attractor subset contains solutions of
observational interest, namely those that have sufficiently large $\Omega_\varphi$ and
accelerated expansion during the last few billion years up to the present time
(see the discussion in~\cite{Alam2003}).
Since these solutions satisfy $V(\varphi)>0$, $\rho_\varphi\geq 0$, $\Omega_\varphi \geq 0$,
from their asymptotic past to the present time, and since thereby $v = \sqrt{\Omega_\varphi/3}$
is well defined for this time period we can use the previous formulation for bounded $\lambda(\varphi)$
for their evolution to the present time. Figure~\ref{fig:thaw2statespace} depicts three
thawing attractor solutions in the two state spaces $(\bar{\varphi},u,v)$ and
$(\bar{s},\bar{\Sigma}_\varphi,\theta)$ from their asymptotic past at $\mathrm{FL}^+$
to the present time (characterized by $\Omega_\varphi = 0.68$ and denoted by a ring
in Figure~\ref{fig:thaw2statespace}) and beyond, where only the
$(\bar{s},\bar{\Sigma}_\varphi,\theta)$ state space
yields their entire evolution and their final contraction to a big crunch at the
kinaton fixed point $\mathrm{K}_+^-$.
%

%
%



\subsection*{Acknowledgments}
A. A. is supported by FCT/Portugal through \\ FCT/Tenure 2023.15700.TENURE.003/CP00055/CT00003, CAMGSD, IST-ID, grant No. UID/4459/2025, and projects 2024.04456.CERN, and  H2020-MSCA-2022-SE EinsteinWaves, GA No. 101131233.
A.A. would also like to thank the CMA-UBI in Covilh\~a for kind hospitality.
C. U. would like to thank the CAMGSD, Instituto Superior T\'ecnico in Lisbon for kind hospitality.

\bibliographystyle{unsrt}
\bibliography{../Bibtex/cos_pert_papers}

\end{document}